\documentstyle{article}

\begin{document}

\begin{center}
{\huge\bf Quantum Hall Effect and Chaotic Motion in Phase Space}
\end{center}

\vspace{1cm}
\begin{center}
{\large\bf
F.GHABOUSSI}\\
\end{center}

\begin{center}
\begin{minipage}{8cm}
Department of Physics, University of Konstanz\\
P.O. Box 5560, D 78434 Konstanz, Germany\\
E-mail: ghabousi@kaluza.physik.uni-konstanz.de
\end{minipage}
\end{center}

\vspace{1cm}

\begin{center}
{\large{\bf Abstract}}
\end{center}

\begin{center}
\begin{minipage}{12cm}
We discuss the relation between the Quantum Hall behaviour of  
charged carriers and their chaotic motion in phase space. It is  
shown that the quantum Hall diagram is comparable with the stepped  
diagram in phase space of a chaotic motion.
\end{minipage}
\end{center}

\newpage

Recent developments in experimental and theoretical modeling of  
quantum Hall effects (QHE) show that charge carriers should have  
chaotic behaviour during their participation in QHE \cite{weis}.

We discuss here the relation between QHE and chaotic motion and  
show that QHE and specially its localization-delocalization  
phenomenology \cite{all} should be considered as a kind of chaotic  
motion in the phase space of system.

Let us first determine what we understand as an appropriate model  
of chaotic motion to be compared with the QHE. It is the so called  
snapshots model which contains as a limiting case the so called  
"kicked" models as well \cite{balaz}. It describes a classical  
motion in the phase space according to a Hamiltonian which is  
alternatively purely kinetik, or purely potential. The motion in  
phase spase is a stepped motion for finite time intervalls  
consisting of alternatively parallel lines to the momentum axis for  
the pure potential Hamiltonian and lines parallel to the position  
axis for purely kinetic motion \cite{balaz}.

Geometrically, such a potential behaviour is equivalent to a  
potential which is limited to differ from zero only in parts of a  
multiply connected region, where this one could be in general a  
time- or a space manifold. A quantization of the related motion then  
requires, in view of the well known flux quantization that this  
potential should be a pure gauge potential \cite{flusme}. Thus, we  
have a potential which exists and disapears frequently in finite  
amount of time or in finite regions of spase, i. e. in a mutiply  
connected region.

Now the importent point is that we have exactly this same potential  
behaviour in the QHE due to the disturbance and perturbations  
caused by impurities, so that the well known diagram of quantized  
Hall- condutivity or resistivity with plateaus is almost of the  
above mentioned stepped form. Therefore, we claim and show that the  
QHE can be considered as a realization of the "quanized" chaotic  
behaviour of the mentioned kind.

Coming back to the description of the mentioned chaotic motion let  
us remark that, in view of its well known non-regular behaviour, it  
can be considered partly as a "finite" limit of the usual classical  
continuous motion in phase space; or equivalently the usual  
classical regular motion can be considered as a continuum limit of  
the chaotic "finite" motion. The finiteness or the partial absence  
of derivative in chaotic motion can appear either with respect to  
the time- or with respect to the position variable depending on the  
experimental set up or on the theoretical model which describes the  
resulting chaotic motion. Thereby, in view of the absence of  
derivatives at any point, the chaotic motion should be given by

\begin{eqnarray}
\Delta q &=& \frac{p}{\mu} \Delta t \nonumber\\
\Delta p &=& - \frac{\Delta V}{\Delta q} \Delta t  \;\;\;,
\end{eqnarray}
\label{erst}

for finite quantities ${\{\Delta q, \Delta p, \Delta V, \Delta  
t}\}$ with $\mu$ representing the mass of the system, where the  
first equation describes the purely kinetik part of motion and the  
second one describes the purely potential part. They can be  
rewritten for differentiable potential $V$ but for finite $\Delta t  
= T$ and $\Delta p = p_{n+1} -p_n$ in the following form which is  
more familiar from Chaos theory \cite{balaz}:

\begin{eqnarray}
(a) \;\;\; q_{n+1} - q_n &=& \frac{p_n}{\mu} T \nonumber\\
(b) \;\;\;  p_{n+1} - p_n &=& - \frac{\partial V}{\partial q} T \;\;\;,
\end{eqnarray}
\label{zweit}

where we used the definitions $\Delta r := r_{n+1} -r_n$.

It is this finiteness of chaotic motion with finite quantities from  
the set ${\{\Delta q, \Delta p, \Delta E, \Delta t}\}$ which is  
closely related to the quantum mechanics, if we recall that  
according to the uncertainty relation we have:

\begin{equation}
\Delta p \cdot \Delta q = \Delta E \cdot \Delta t \geq \hbar
\end{equation}
\label{dritt}

Here also we have to do with finite magnitudes of momentum,  
position etc. which can not vanish \cite{direct}.

The finite quantum structure arises here in view of the fact that  
$\Delta E = E_{n+1} - E_n = E_0 = \displaystyle{\frac{\hbar  
\omega}{2}}$ where $\Delta E$ defines the smallest amount of energy  
which is available quantum mechanically and which can not be  
undercut. Thus, it determines also the smallest quantum mechanical  
length and momentum which are considerable and which can not be  
undercut quantum mechanically. An example of such finite smallest  
quantum length is the well known magnetic length $l_B$ which plays  
an importent role just in the QHE \cite{all}. It is the width of a  
"quantum ring" where the edge currents flow.

Accordingly, if one translates the QHE behaviour into the quantum  
chaotic terminology, then

$\Delta q = q_{n+1} - q_n$ in the QHE case should be given by the  
magnetic length

\begin{equation}
(\Delta q)^2 = l_B^2 = \frac{\hbar}{e B} = \frac{\nu}{2\pi n}\;\;\;,
\end{equation}
\label{viert}

where $e, \nu$ and $n$ are the elementary charge, filling factor  
and carrier concentration respectively.

It is the "unit" length or heigths of the ideal staires, i. e.  
parallel to the position or to the momentum axis in the phase space  
diagram according to the chaos theoretic description of QHE, which  
is given by the relation (4) according to the magnitude of the  
applied exterior magnetic field $B$ or  
$\displaystyle{\frac{\nu}{n}}$ in QHE samples \cite{all}.

Such a translation means on the other hand that the chaotic motion
(2) is quantized according to the usual quantization methodes in  
Chaos theory \cite {Gutz} \cite{balaz}.
Thus, the QHE behaviour can be translated according to the  
finiteness of quantum phase space cells into the mentioned finite  
"quantum" chaotic motion.

After this qualitative remarks we give a potential analysis of the  
localization-delocalization transition which is responsible for the  
typical step structure of the quantized Hall resistivity.

First of all let us mention that according to our understanding of  
IQHE, if one considers the ideal stepped diagram of quantum Hall  
resistivity or conductivity against the applied magnetic field  
strength, the plateaus mark positions where the field strength is  
not present and the vertical parts of diagram mark positions where  
it is present. It is the same situation as for a repeated  
Bohm-Aharonov set ups where the field strength is present in some  
parts but absent in other parts of space, whereby the difference of  
regions in QHE case is due to the impurities. As mentioned bevor,  
geometrically we have to do with a multiply connected region where  
there are "holes" which are comparable qualitatively with bounded  
regions of action of potentials or with dots and antidots structur.  
One can understand the IQHE situation directly from the circumstance  
that a present field strength or a present potential gives rise to  
a change of momentum, velocity or current density, whereas the  
absence of field strength or its potential results in the contancy  
of the current density in view of the Ohm's equations, as in  
relation (2(b)), in accordance with the above discussed chaotic  
motion diagram.

Moreover, according to our gauge field theoretical  
Chern-Simons-Schroedinger model of IQHE, it is the presence of  
electromagnetic potential in the primary version of current density  
relation of the classical Hall effect which results in the term  
proportinal to the longitudinal conductivity \cite{mein}. These  
positions in the diagram of longitudinal conductivity are according  
to the well known joint diagrams of Hall- and longitudinal  
conductivities in IQHE \cite{all} exactly the positions where the  
Hall conductivity is in vertical part of its digram, i. e. in its  
purely potential domain according to the above discussed chaos  
terminology \cite{neu}. On the other hand, in the plateau regions of  
Hall conductivity the longitudinal conductivity is almost vanishing  
\cite{almost}, which fits also in the terminology of the chaos  
theory, according to which these positions are the purely kinetic  
positions.

Thus, the integer quantum Hall conductivity diagram describes a  
motion in its own phase space which is similar to the motion  
described by the diagram of chaotic motion in the above discussed  
phase space, where in both cases the potential is alternatively  
present and absent in the Hamiltonian which results in the purely  
potential and purely kinetic motion respectively \cite{balaz}. To be  
complet, let us mention that in both cases there are the purely  
potential part of motion which causes the change in momentum  
resulting in lines which are parallel to the momentum axix in  
diagram. Moreover, in both cases there are the purely kinetic part  
which causes the change in the position resulting in lines which are  
parallel to the position axix or to the axix which represents the  
conjugate variable in phase space as a function of position  
variable, e. g.  potential or field strength.

Furthermore, if the potential becomes in the kinetic domain  
absolutely zero, then we have the exact parallel lines to the  
position axis, whereas in quantum case the lines can become almost  
but never exactly parallel in view of the mentioned quantum  
residuens. Accordingly, in quantum mechanical cases like QHE we have  
always only almost purely potential or almost purely kinetic  
Hamiltionian or motion \cite{almost} and the diagrams are also  
modified accordingly.

\medskip
Of course the phase space of the IQHE system should be considered,  
also according to our model \cite{mein}, electrodynamically, i. e.  
with potential components as phase space variabels \cite{wittme}.  
Nevertheless, in view of the invariance of the action functional  
with respect to the canonical transformations in phase space one  
should red up the phase space variables also from actions like $  
e\oint A_m dx^m = e\int\int B ds$ with $m = 1,2$ and $s = area$,  
where $A_m$ is the electromagnetic potential and $B$ is the magnetic  
field strength.
Thereafter, the mentioned stepped diagram of IQHE can be  
considered, in view of the Ohm's equations $ne  
\displaystyle{\frac{dx}{dt}} = j_m = \epsilon_{nm}\sigma_H E_n =  
\epsilon_{nm}\sigma_H \displaystyle{\frac{dA_n}{dt}}$ with  
$\epsilon_{mn} =
- \epsilon_{nm} = 1$, as a chaotic motion between $x:=q$ and $A:=p$  
variables which are related by $\sigma_H$ in the phase space of  
IQHE or between other equivalent canonically conjugate phase spaces  
variables. An example of such equivalent set of phase space  
variables is ${\{\sigma_H [{\
\displaystyle{\frac{e^2}{h}}}] \;, \nu}\}$ where the diagram in  
phase space, i. e. $\sigma_H$ against $\nu$ is of exact step form,  
whereby the joint diagram of longitudinal conductivity consists of  
vertical lines on the $\nu$ axis [2e].

We mentioned already that from our gauge theoretical point of view  
\cite{mein} the IQHE diagrams could be understood according to the  
repeated Bohm-Aharonov effects \cite{all} on the IQHE sample, i.e.  
in view of the presence and absence of potential or its field  
strength in different space regions, where this presence and absence  
depend on the constitution of region according to the distribution  
of the impurities. In other words, a simple model can be thought so  
that the potential is non-vanishing only in the impurity regions,  
whereas it is almost vanishing in other parts of sample  
\cite{almost}. However, its line integrals $\oint A$ in these last  
parts of region is not zero \cite{flusme}, as it is also known from  
the Bohm-Aharonov effect. Moreover, the value of integral should  
increase with the length of circumference of the closed path which  
is related with the number of antidots inside the path  
\cite{almost}. Such a behaviour is, as already mentioned, the  
typical behaviour of an almost pure U(1) or electromagnetic gauge  
potential \cite {flusme} \cite{almost}.

It is intresting to mention that the variation of the action  
functinal $\oint (p - eA)$, which is sugessted in Ref. \cite{weis}  
to explain the chaotic behaviour in IQHE, has the same almost pure  
electromagnetic gauge potential as its solution.

Furthermore, the quantum oscillations of longitudinal resistivity  
in the $B \rightarrow 0$ region reported in Ref.\cite{weis} which  
appear in "very low
temprature" should arise just due to the quantum mechanical  
uncertainty relation between energy and time in the following  
manner, if we recall the above analysis of chaotic motion:

In the case of chaotic motion, if the time $T$ tends to zero, i. e.  
$T \rightarrow 0$, then the seccession of the horizontal and  
vertical lines approximates more and more the usual continous  
trajectory which is associated with the regular continuous motion in  
the phase space arising from a conserved Hamiltonian \cite{balaz}.  
However, if $T$ is finite, then the staires in the mentioned diagram  
become revealed in view of the fact that the pure potential and  
pure kinetic periods are remarkable.
On the other hand, according to the uncertainty relation $\Delta E  
\cdot \Delta t = \hbar$, very low energy $\Delta E_{\ll}$ or very  
low temprature is related with a non-vanishing and relatively  
remarkable $ T = \Delta t = \displaystyle {\frac{\hbar}{\Delta  
E_{\ll}}}$. Thus, in view of our discussion the staires which were  
negligable in the $ T\rightarrow 0$ limit become revealed in the new  
situation. Accordingly, the longitudinal conductivity coming from  
purely potential positions in the step, i. e. in the vertical lines,  
becomes also revealed in the same manner as it is discussed above  
\cite{mein}. Recall furthermore that, in the ground state of QHE,  
very low energy is related with very low magnetic field $B$  
according to $\Delta E = E_0 = \displaystyle{\frac{\hbar e B}{2  
\mu}}$, thus at very low magnetic field a very low energy implies  
non-vanishing and remarkable period of time where the potential acts  
bevor it becomes almost vanishing during the next finite time and  
so the staires and the related resonances become revealed (see also  
Ref. \cite{balaz}).
Moreover, such a mechanism could be modeled in the way that the  
antidots become related with impurities or holes in the repeated  
Bohm-Aharonov context according to the following consideration:

Although $B$ is very low in the region under consideration, however  
the magnetic flux $\int\int B ds = \oint A$ becomes larger, if the  
electronic orbits surround bigger areas, i. e. with more antidots.  
In view of the fact that not $B$ but only invariants like $\int\int  
B ds = \oint A$ are relevant in quantum mechanical set ups as in  
QHE, one measures in QHE quantities which depend not only on $B$ but  
on the flux of $B$, as it is reported in Ref. \cite{weis}. Thus,  
the smallness of the $B$ value can be compensated by the largeness  
of the area in $\int\int B$ and results in the revelead resonances  
\cite{weis}.

Furthermore, let us mention that the potentials which are used in  
Refs. \cite{weis} \cite{fleish} to model the hard wall billiard and  
the antidots respectively, are all variations of a U(1) (or  
electromagnetic) pure gauge potential in a multiply connected region
as it is obvious from the action used in Ref. \cite{weis} mentioned  
above. It is the non-trivial homotopy of such a multiply connected  
$2-D$ region which is also reflected by the non-trivial holonomy of  
the $U(1)$ bundle over such a region, which becomes manifested by  
the quantum effects on the mentioned $2-D$ region and which can be  
described consistently only according to quantum mechanics.

\medskip
It should be mention at last that just the use of KAM-theorem in  
QHE case in Ref. \cite{fleish} is a hint in direction of a  
perturbation which is a quantum mechanical relict and not a true  
classical mechanical matter. We discuss this question in view of its  
general application seperately \cite{under}.

Footnotes and references


\begin{thebibliography}{100}
\bibitem{flusme}
Considering the usual flux quantization according to $\oint A =  
\int\int B ds = \displaystyle{\frac{h}{e}}\Phi$ with $\Phi =2\pi N  
\;\;, \; N \in {\bf Z}$ it is obvious that the potential is given by  
$A_m = \partial_m \Phi$ (see also Ref. [2]) with $A:=A_mdx^m$. Of  
course, the $F_{mn}(A_m) = B$ is outsides of the area $s$ zero as in  
the Bohm-Aharonov case. Newertheless, in a multiply connected  
region one has $\oint A \neq 0$, as it is known also from the  
Bohm-Aharonov case.

Furthermore, recall that 1) one should consider the multiply  
connected region in phase space which is, in general, not the same  
as the position space and 2) that the multiply connectedness of a  
region is related with the mutiply valuedness of fuctions defined  
thereon.
\bibitem{direct}
Recall also that one can extract from the relation (3) directly  
part of the above mentioned relations (1) for chaotic motion, if one  
consider $\mu \Delta E = p \Delta p := \displaystyle{\frac{1}{2}}  
\Delta (p^2)$.
\bibitem{almost}
One must take into account that despite of classical physics in  
quantum physics one has in view of the uncertainty relations finite  
quantities, e. g. in ground states, which prevent related relevant  
quantities to become absolutely zero. To these relevant quantities  
in the QHE case it belongs the electromagnetic potential or its  
field strength. Equivalently, in quantum mechanics only global  
quantities $\oint A = \int\int B ds$ are relevant but not their  
local components $A$ or $B$. For a constant non-vanishing value of a  
pure gauge potential the value of its closed path integral becomes  
proportinal to the length of the circumference of the path.

Moreover, also the question of absolute zero temperature is related  
with this same quantum mechanical ground state circumestance.
\bibitem{neu}
Recall also that, in accordance with the discussed digram of  
chaotic motion, the field strength $B = F_{mn}$ in the vertical  
parts of diagram have constant values as it is obvious in a diagram  
with $B$ axis as its horizontal axis.
\end{thebibliography}
\end{document}